\documentclass[pra,twocolumn,superscriptaddress,showpacs,longbibliography,a4paper]{revtex4-1}
\usepackage{mathrsfs,braket}
\usepackage{amssymb, amsbsy, amsmath, latexsym, dsfont, array, layout, graphicx, mathrsfs,color, bm, bbm}
\usepackage{makecell,multirow}
\usepackage[colorlinks,linkcolor=blue,anchorcolor=red,citecolor=red]{hyperref}

\newcommand{\one}{\mathds{1}}

\newcommand{\red}{\color[rgb]{0.8,0,0}}

\begin{document}

\title{Experimental realization of a quantum image classifier via tensor-network-based machine learning}

\author{Kunkun Wang}
\affiliation{Beijing Computational Science Research Center, Beijing 100084, China}
\affiliation{School of Physics and Optoelectronics Engineering, Anhui University, Hefei 230601, China}
\author{Lei Xiao}
\affiliation{Beijing Computational Science Research Center, Beijing 100084, China}
%\author{Barry C. Sanders}
%\affiliation{Institute for Quantum Science and Technology, University of Calgary, Alberta T2N 1N4, Canada}
\author{Wei Yi}\email{wyiz@ustc.edu.cn}
\affiliation{CAS Key Laboratory of Quantum Information, University of Science and Technology of China, Hefei 230026, China}
\affiliation{CAS Center For Excellence in Quantum Information and Quantum Physics, Hefei 230026, China}
\author{Shi-Ju Ran}\email{sjran@cnu.edu.cn}
\affiliation{Department of Physics, Capital Normal University, Beijing 100048, China}
\author{Peng Xue}\email{gnep.eux@gmail.com}
\affiliation{Beijing Computational Science Research Center, Beijing 100084, China}

\begin{abstract}
Quantum machine learning aspires to overcome intractability that currently limits its applicability to practical problems. However, quantum machine learning itself is limited by low effective dimensions achievable in state-of-the-art experiments. Here we demonstrate highly successful classifications of real-life images using photonic qubits, combining a quantum tensor-network representation of hand-written digits and entanglement-based optimization. Specifically, we focus on binary classification for hand-written zeroes and ones, whose features are cast into the tensor-network representation, further reduced by optimization based on entanglement entropy and encoded into two-qubit photonic states. We then demonstrate image classification with a high success rate exceeding $98\%$, through successive gate operations and projective measurements. Although we work with photons, our approach is amenable to other physical realizations such as nitrogen-vacancy centers, nuclear spins and trapped ions, and our scheme can be scaled to efficient multi-qubit encodings of features in the tensor-product representation, thereby setting the stage for quantum-enhanced multi-class classification.
\end{abstract}

\maketitle

\section{Introduction}
\label{sec:intro}

The interdisciplinary field of quantum machine learning (ML) has seen astonishing progress recently~\cite{BWPR+17QML,DB18}, where novel algorithms presage useful applications for near-term quantum computers. A concrete example is pattern recognition, where accurate modeling requires an exponentially large Hilbert-space dimension, especially for quantum classifiers which can lead to unique advantages over their classical counterparts~\cite{HCT+19QML, CCL19QCNN, MN19QML}.
Such a quantum advantage derives from the efficient exploitation of quantum entanglement that also underlies the extraordinary interpretability of tensor networks (TNs), a powerful theoretical framework originating from quantum information science and with wide applications in the study of strongly-correlated many-body systems~\cite{VMC08MPSPEPSRev, CV09TNSRev, HV17TMTNrev, RTPC+17TNrev}.
Recent works on TN-based ML algorithms, due to their quantum nature, exhibit competitive, if not better, performance compared to classical ML models such as supportive vector machines~\cite{HCT+19QML, SPLRS19GTNC, RML14SVM, LLXD15SVMNMR, ARR03SVM} and neural networks~\cite{YYQ18RTNML, LSCS19EntML, DLD17NNent,SS16TNML, LRWP+17MLTN, HWFWZ17MPSML, GJLP18MPOML, S18MERAML, CWXZ19generateTTNML, GSPEC19PTNML, EHL19MPSML}.
It is thus tempting to demonstrate TN-based ML algorithms in genuine quantum systems, with the hope of tackling practical tasks. However, major obstacles exist, as TN-based ML algorithms typically require an unwieldily large Hilbert-space dimension to process real-life data. The problem is made worse by the limited number of noisy qubits on currently available quantum platforms. So far, TN-based ML has yet to be demonstrated on any physical system.

In this work, we demonstrate TN-based ML schemes with single photons for the first time, and apply the scheme to a popular problem---optical character recognition, to classify real-life, hand-drawn images. As a key element of our scheme, we combine the interpretability of TN with an entanglement-based optimization~\cite{LZLR18entTNML}, such that the dimension of the required Hilbert space is dramatically reduced. We are then able to implement the corresponding qubit-efficient quantum circuits~\cite{HPWS18TNQML} through single-photon interferometry.

Focusing on the minimal task of binary classification of hand-written digits of ``$0$'' and ``$1$''~\cite{MNIST}, we demonstrate two TN-based ML schemes (A and B), each with three- and five-layer constructions, corresponding to the dimension of the quantum feature space. The gate operations for the classifier are trained and optimized through supervised learning on classical computers, and results of the classification are read out through projective measurements on the output photons. Experimentally, we achieve an over $98\%$ success rate with both of our schemes for classifying all testing images of ``$0$'' and ``$1$'' in the MNIST dataset~\cite{MNIST}.

We further demonstrate exemplary cases where the post-training classifier correctly recognizes poorly written digits that are not in the MNIST dataset, thus confirming the robustness of our classifier.
%While our hybrid quantum-classical optimization scheme can be further upgraded to be fully quantum mechanical, the TN-based ML algorithm demonstrated here is quite general and scalable.
Together with recent progress of ML either on quantum systems~\cite{CWSC+15MLQCP,LLXD15SVMNMR,ZZW20QMLion,AABB+2019GoogleQ,HCT+19QML,PTP19,LPSW17,GZD2020}, or with classical-quantum hybrid setups~\cite{ZLBL+19HQML,HPWS18TNQML,BGC+19,GQJ+18,HWC+19,HS10,LWL+19}, our experiment paves the way for quantum advantages in solving real-world problems.

%------------------------------------------------------------------------------------%

\section{Supervised machine learning by tensor network}

One of the main challenges in dealing with real-life data using quantum devices is the requirement of large numbers of qubits ($\sim O(10^2)$ or more)~\cite{XUE06,XUE09}. To address the issue, we apply an entanglement-based feature extraction~\cite{LZLR18entTNML,HPWS18TNQML} to implement a TN-based, qubit-efficient image classifier using single photons.

As illustrated in Fig.~\ref{fig:fig1}, our scheme breaks down into the following steps (see more details in Appendix A): (i) we map the full data of classical images to quantum states, and train the matrix product state (MPS) classifier using a supervised TN-based ML algorithm with $N$ features (pixels or frequency components; $N=784$ for each image in the MNIST). Here we consider a binary classification problem to classify hand-written digits of ``$0$'' and ``$1$'' in the MNIST dataset into $N_c=2$ categories.
In step (ii), we extract a handful of the most important features using an entanglement-based optimization. In step (iii), a new MPS is constructed and then trained with the extracted features obtained in step (ii). A small number of feature qubits ($\tilde{N}=3$ or $5$ feature qubits corresponding to three- or five-layer constructions of the classifier) with the largest entanglement entropy in the quantum feature space are retained~\cite{LZLR18entTNML}, to represent and classify hand-written digits of ``$0$'' and ``$1$'' in the MNIST dataset.
These three steps correspond to the classical optimization and feature extraction, shown in Fig.~\ref{fig:fig1}.

\begin{figure}
\includegraphics[width=0.45\textwidth]{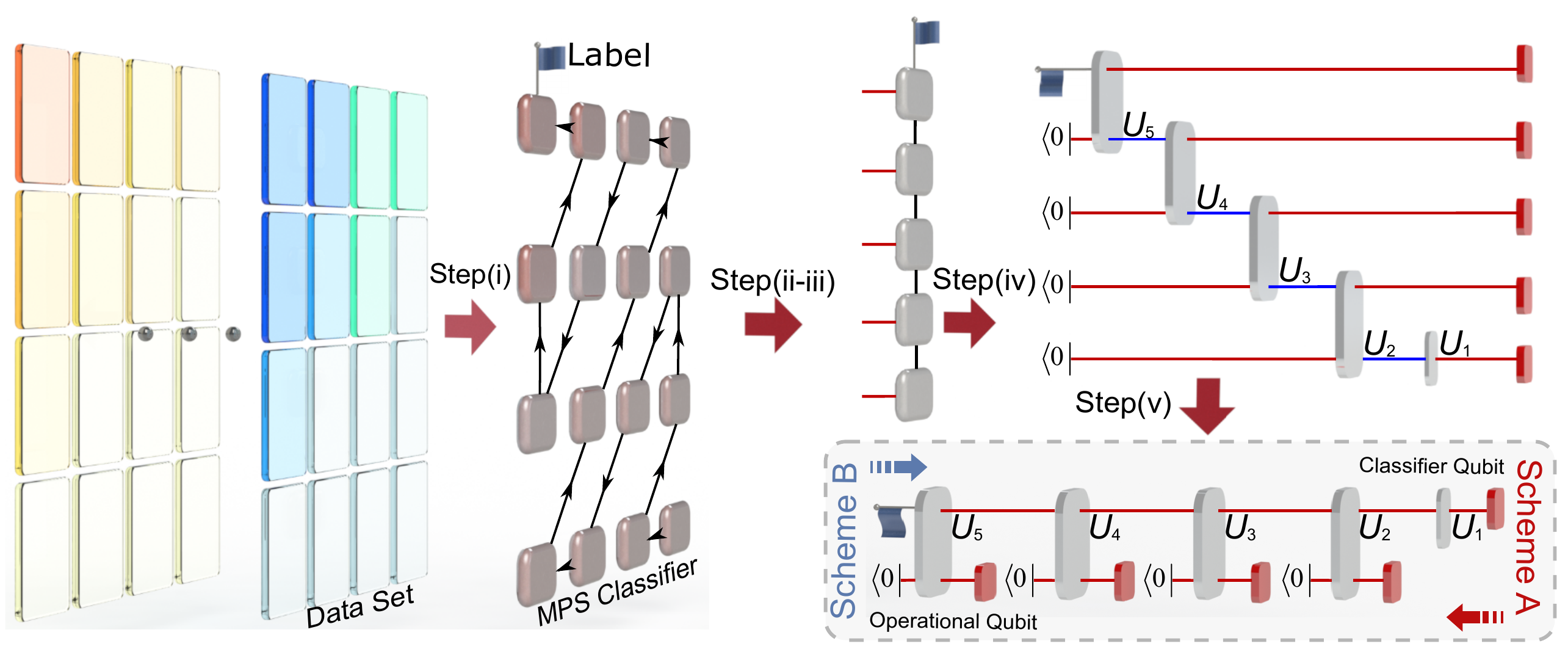}
\centering
\caption{Illustration of the main steps of implementing the TN-based quantum classifier. In step (i)-(iii), we map the images of $N=784$ pixels (features) to quantum states, and train the matrix product state (MPS) classifier using these images in the MNIST dataset by a supervised learning algorithm. The MPS consists of $N$ tensors represented by squares, with the label index represented by a flag. We calculate the entanglement entropies of the MPS and retain only $\tilde{N}=3$ or $5$ features with the largest entanglement entropies. Then we train a reduced MPS with the $\tilde{N}$ retained features. In step (iv)-(v), the reduced MPS is translated to a circuit acting on $\tilde{N}$ qubits, and further to a circuit using the qubit-efficient scheme, which involves $\tilde{N}'=2$ qubits and $\tilde{N}$ gates including a single-qubit gate $U_1$ and $\tilde{N}-1$ two-qubit gates $U_i$ ($i=2,3$ or $i=2,...,5$). We employ two schemes to implement the classifier, whose orders of operations are indicated by the red and blue arrows, respectively.
}
\label{fig:fig1}
\end{figure}

\begin{figure}[t]
\includegraphics[width=0.45\textwidth]{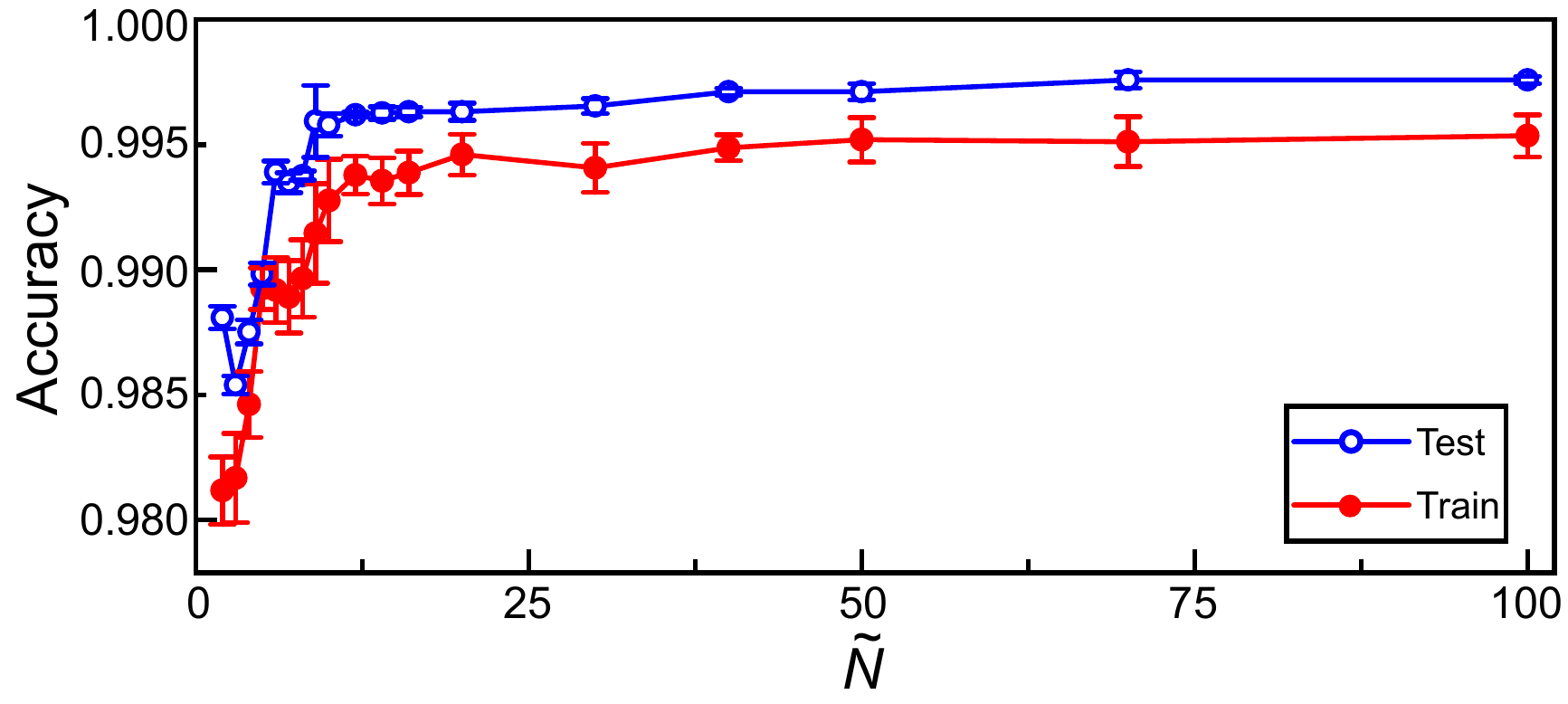}
\centering
\caption{Training and testing accuracies of classifying the ``0'' and ``1'' digits in the MNIST dataset as functions of $\tilde{N}$. Data are obtained by averaging the accuracies over $20$ numerical simulations, and error bars indicate the statistical uncertainty.
}
\label{fig:newfig2}
\end{figure}

In Fig.~\ref{fig:newfig2}, with increasing $\tilde{N}$, the testing accuracy and the training accuracy for the MNIST dataset both increase quickly as expected. More important, the fact that both accuracies converge for $\tilde{N}\geq10$. For $\tilde{N}=3,5$, both accuracies are higher than $0.98$, which implies that the reduced number of the feature qubits works well for the TN-based quantum classifier.

The subsequent experimental implementation involves the following two steps (see more details in Appendix B).
In step (iv), we translate the tensors in the optimized MPS to quantum circuit, which form a circuit acting on $\tilde{N}$ qubits~\cite{SSVCW05PrepareMPS}. In step (v), through the quantum-efficient scheme~\cite{HPWS18TNQML}, the quantum circuit is further simplified to the one with $\tilde{N}$ gates acting on $\tilde{N}'$ qubits. We take $\tilde{N}'=2$, where $2$ qubits are dubbed as the operational and classifier qubits. This is achieved by translating measurements on different qubits into those on only $2$ qubits at different times. The extracted features of an image are input to the circuit by measuring the operational qubit at different times. The output of the image classifier is accessed via projective measurements on the classifier qubit.

We present two schemes of TN-based quantum classifier, each being the reverse of the other.
In either case, we embed core features of an image into the quantum feature space of three or five feature states $\{|\psi_i\rangle\}$, which corresponds to a three- or five-layer construction that involves their respective number of unitary gate operations.

\begin{figure}[tb!]
\includegraphics[width=0.45\textwidth]{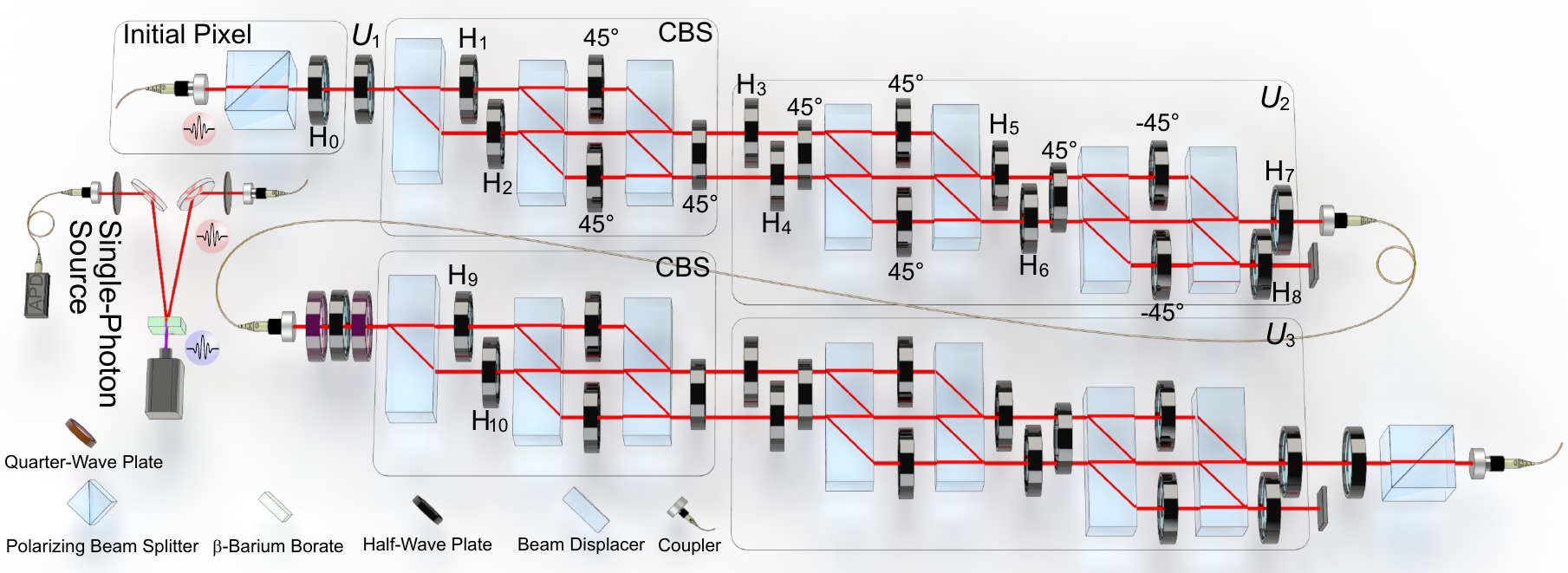}
\centering
\caption{Experimental demonstration of quantum classifier with the three-layer construction.
For each pair of photons generated by spontaneous parametric down conversion, one photon serves as the trigger and the other, the signal photon, proceeds through the experimental setup corresponding to the two schemes.
Under Scheme A, the signal photon is projected onto the polarization state $|\psi_1\rangle$ of the classifier qubit via a PBS and a HWP H$_0$. It is then sent through an interferometric network, composed of HWPs and BDs, before detected by APDs in coincidence with the trigger photon. The single-qubit gate $U_1$ is realized by a HWP, and two-qubit gates $U_2$ and $U_3$ are realized by BDs and HWPs. The input states $|\psi_2\rangle$ and $|\psi_3\rangle$ are encoded in the spatial modes of photons, and are prepared via the HWPs (H$_1$, H$_2$, H$_9$, and H$_{10}$) in controlled beam splitters (CBSs) which, by combining the spatial and polarization degrees of freedom, effectively expand the dimension of the system.
A $1$m long single-mode fiber serves as a spatial filter in between successive modules of two-qubit gate, and a set of wave plates is introduced to offset the impact of the fiber on the photon polarizations. For detection, photons are projected onto upper spatial mode $|u\rangle$ by discarding those in the lower one, and a projective measurement of $\sigma_z$ is realized through a HWP, a PBS and APDs. Scheme B is the exact reverse process of Scheme A, where photons are sent into the setup through the output port of Scheme A.
}
\label{fig:fig2}
\end{figure}

In Scheme A, we initialize the classifier and operational qubits into feature states $|\psi_1\rangle$ and $|\psi_2\rangle$, respectively. Other feature states $|\psi_i\rangle$ are used as successive inputs for the optimized quantum circuit consisting of a series of single- and two-qubit gates $U_i$. Taking the three-layer construction as an example, the classifier qubit is subject to the single-qubit operator $U_1$, whose output is fed into the two-qubit operation $U_2$ together with the operational qubit. After projecting the operational qubit into the basis state $|0\rangle$, the classifier qubit is characterized by the density matrix
\begin{equation}
\rho_1=\text{Tr}_{\rm op}\left[U_2\left(U_1|\psi_1\rangle\langle\psi_1|U^\dagger_1\otimes|\psi_2\rangle\langle\psi_2|\right)U^\dagger_2\left(\one\otimes|0\rangle\langle0|\right)\right].\notag
\end{equation}
Here $\text{Tr}_{\rm op}$ is the trace over the two-dimensional Hilbert space of the operational qubit. The operational qubit is then prepared in the feature state $|\psi_3\rangle$, before the two-qubit operator $U_3$ is applied on both qubits. After projecting the operational qubit into the basis state $|0\rangle$ again, the classifier qubit is given by
\begin{equation}
\rho_2=\text{Tr}_{\rm op}\left[U_3\left(\rho_1\otimes|\psi_3\rangle\langle\psi_3|\right)U^\dagger_3\left(\one\otimes|0\rangle\langle0|\right)\right].\notag
\end{equation}
We perform a projective measurement of $\sigma_z=\ket{0}\bra{0}-\ket{1}\bra{1}$ on the output classifier qubit, yielding probabilities $P_0$ and $P_1$. The image is recognized as ``$0$'' (``$1$'') for $P_0>P_1$ ($P_1>P_0$).

In Scheme B, we initialize the classifier and operational qubits into the two-qubit state $|00\rangle$ (or $|10\rangle$), and
successively apply the optimized unitary gates $U_i^{\dagger}$ (in the reverse order compared to that in Scheme A). A projective measurement $|\psi_i\rangle\langle\psi_i|$ is performed following the corresponding unitary operation $U_i^{\dagger}$. The last projective measurement $|\psi_1\rangle\langle\psi_1|$ on the classifier qubit yields the probability $P_0'$ ($P_1'$) for the initial state $|00\rangle$ (or $|10\rangle$). Since $P_0=P_0'$ and $P_1=P_1'$ (see the derivations in Appendix C), the image is recognized as ``$0$'' (``$1$'') for $P_0'>P_1'$ ($P_1'>P_0'$).

\begin{figure}
\includegraphics[width=0.45\textwidth]{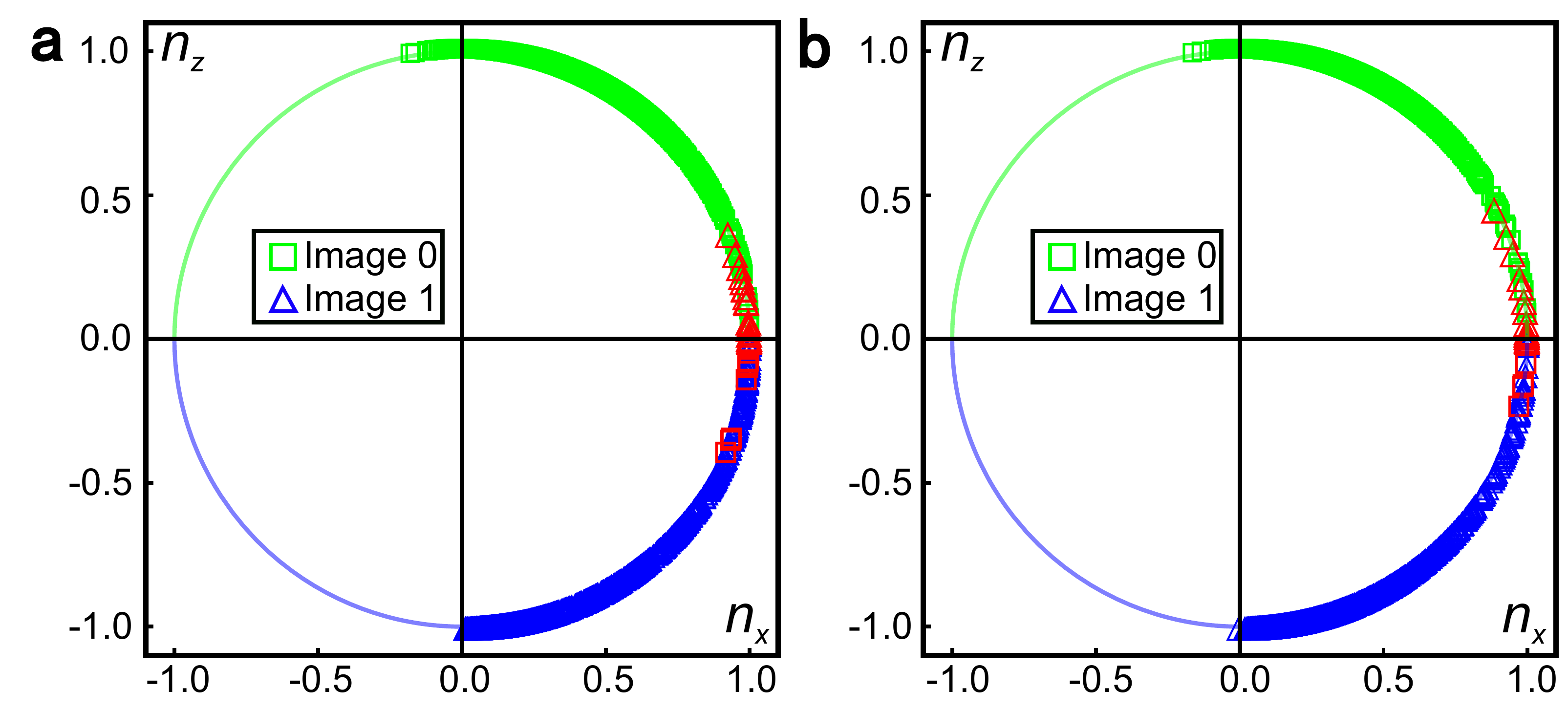}
\centering
\caption{Theoretical results of the testing set under the three- a and five-layer b constructions, respectively. The output states are shown in the $x$-$z$ plane of Bloch sphere associated with the classifier qubit. The green squares (blue
triangles) indicate successful recognition of the hand-written digit ``$0$'' (``$1$''). Red symbols indicate cases in which the classifier fails to recognize the image.
}
\label{fig:fig3}
\end{figure}

\begin{figure*}
\centering
\includegraphics[width=0.8\textwidth]{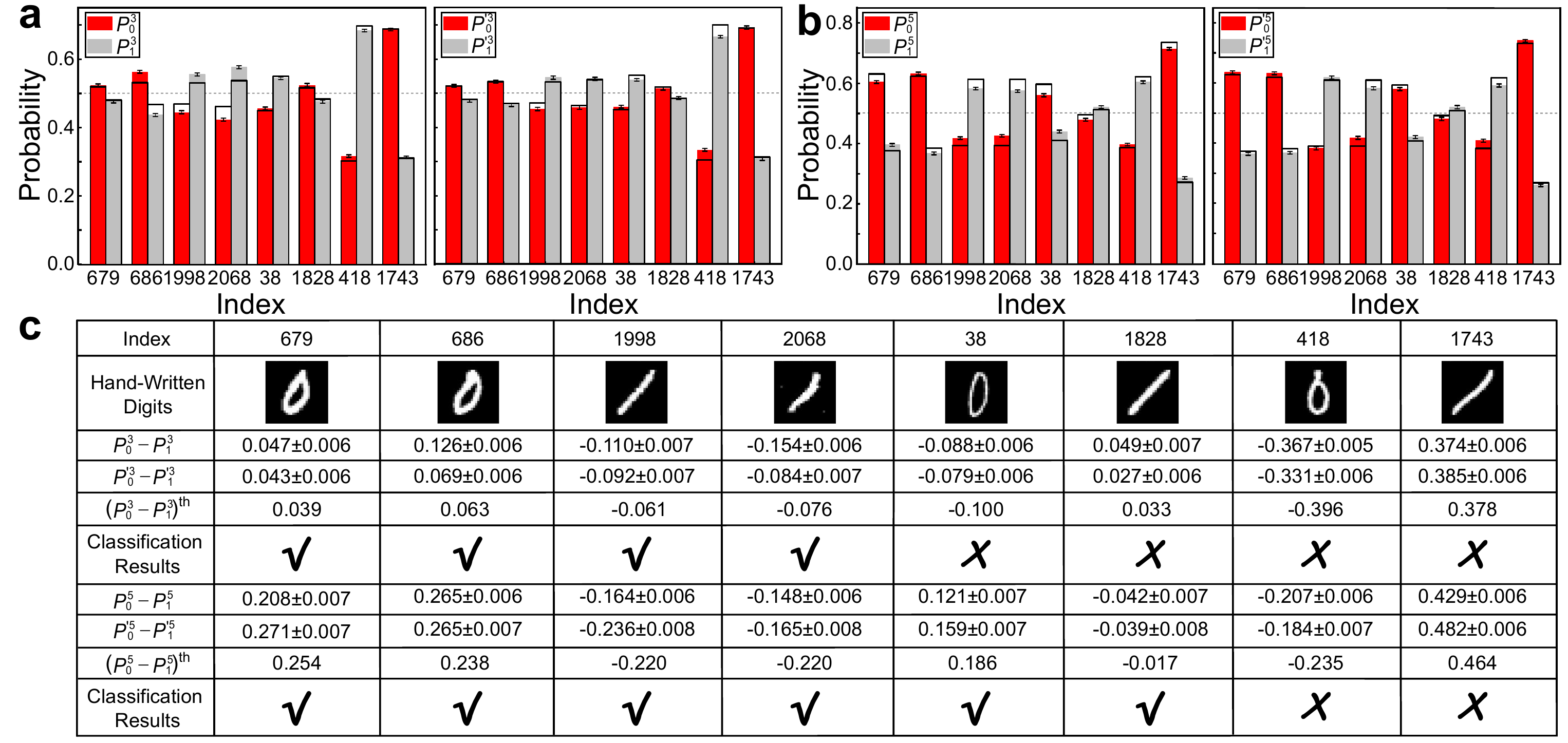}
\centering
\caption{Experimental classification of images within the testing set. Measured probabilities of the projective measurements on output states of quantum classifiers with the three- (a) and five-layer (b) constructions. Left: Scheme A; right: Scheme B. Colored bars represent experimental results, while hollow bars represent their theoretical predictions. Error bars indicate the statistical uncertainty, obtained by assuming Poissonian statistics in the photon-number fluctuations. (c) The classification results for eight typical hand-written digits in the testing set. Rows represent the index of images, namely the hand-written digits, the experimental and theoretical probability differences, and classification results. $P^3_{0,1}$ and $P^5_{0,1}$ represent probabilities  associated with the three- and five-layer constructions, respectively.
}
\label{fig:fig4}
\end{figure*}
%------------------------------------------------------------------------------------%
\section{Experimental realization}
Experimentally, we encode the classifier qubit in the polarization states of the signal photons, i.e., $|H\rangle=|0\rangle$ and $|V\rangle=|1\rangle$ (see Fig.~\ref{fig:fig2}). The operational qubit is encoded in the spatial modes of the photons, with $|u\rangle=|0\rangle$ and $|d\rangle=|1\rangle$ representing the upper and lower spatial modes, respectively. While the single-qubit gate $U_1$ is implemented using a half-wave plate (HWP), the two-qubit gates $U_i$ ($i=2,3$) are implemented through cascaded interferometers, consisting of HWPs and beam displacers (BDs). In Scheme A, feature states $|\psi_i\rangle$ ($i=2,3$) are introduced via a controlled beam splitter (CBS), which consists of HWPs and BDs, with information of the feature states encoded in the setting angles of the HWPs.

The gates are trained by $12665$ hand-written digits of ``$0$'' and ``$1$'' in the training set of the MNIST dataset~\cite{MNIST}. After the training process, $U_i$ are fixed for subsequent image recognition. To assess the training process, we first use $2115$ images in the testing set of the MNIST dataset as input. Specifically, we use all the training images of ``$0$'' and ``$1$'' for training and all testing images for testing in the MNIST dataset, which are all downloaded from the official website. As illustrated in Fig.~\ref{fig:fig3}, under the three-layer construction, the classifier fails to recognize only $30$ out of the $2115$ images, including $10$ images of ``$0$'' and $20$ images of ``$1$''. The success probability of classification is $0.9858$. Under the five-layer construction, the classifier fails to recognize $19$ images within the same testing set, including $4$ images of ``$0$'' and $15$  images of ``$1$''. The success probability is $0.9910$.

Figure~\ref{fig:fig4} demonstrates in details the results of several typical testing images as examples. For all chosen images,
the experimental results suggest that the classifiers are well-trained, in the sense that their predictions have a high success rate, even if the probability difference $P_0-P_1$ ($P'_0-P'_1$) can be small for certain cases. Furthermore, some images that cannot be classified under the three-layer construction can be successfully classified under the five-layer construction, confirming the improvement in behavior of the quantum classifier with an increased feature-space dimension.

To estimate the deviation of the experimental results and theoretical predictions, we define a distance
\begin{equation}
d:=\sqrt{\frac{(P_0-P_0^\text{th})^2+(P_1-P_1^\text{th})^2}{2}},\notag
\end{equation}
where $P_{0,1}$ and $P_{0,1}^\text{th}$ are the measured probabilities and theoretical probabilities, respectively. The distance varies between $0$ and $1$, with $0$ indicating a perfect match. Of all eight distances for the images in Fig.~\ref{fig:fig4}, the largest is $0.039\pm0.003$, which indicates that our experimental results are in excellent agreement with theoretical predictions.

The differences between the experimental data and theoretical predictions are caused by several factors, including fluctuations in photon numbers, the inaccuracy of wave plates, and the dephasing due to the misalignment of the BDs. To provide a quantitative estimate of the success rate of our quantum classifier, we perform numerical simulations by taking experimental imperfections into account.

First, the imperfection caused by photon-number fluctuations increases with decreasing photon counts. In our experiment, the total photon count for each image classification is larger than $10^4$. Therefore, we adopt a total photon count $10^4$ for our estimation, while assuming a Poissonian distribution in the photon statistics. Second, parameters characterizing the inaccuracy of wave plates and the dephasing are estimated using experimental data in Fig.~\ref{fig:fig4}. Specifically, for each wave plate, we assume an uncertainty in the setting angle $\theta+\delta\theta$, where $\delta\theta$ is randomly chosen from the interval $\left[-1.588^\circ,1.588^\circ\right]$ ($\left[-1.180^\circ,1.180^\circ\right]$) under the three-layer (five-layer) construction. The range of these intervals are determined through Monte Carlo simulations, to fit the deviations of experimental data from their theoretical predictions for the eight images in Fig.~\ref{fig:fig4}.
On the other hand, the dephasing due to the misalignment of BDs affects the experimental results through a noisy channel $\varepsilon(\rho)=\eta \rho+(1-\eta)\sigma_z\rho\sigma_z$ characterized by a dephasing rate $\eta$~\cite{WWZ+18}, where $\rho$ ($\varepsilon(\rho)$) is the density matrix of the input (output) state of the noisy channel. By numerically minimizing the difference between the numerical results and the corresponding experiment data, we estimate $\eta$ to be $0.9977$ ($0.9926$) for three-layer (five-layer) construction.

With these, we perform Monte Carlo simulations of our experiments on all $2115$ images in the testing set, from which a success rate is estimated. We then repeat the process for $100$ times and keep the lowest success rate as our final estimation. The estimated success rates are the following: $0.9825$ (Scheme A with three-layer construction); $0.9820$ (Scheme B with three-layer construction); $0.9877$ (Scheme A with five-layer construction); $0.9877$ (Scheme B with five-layer construction). Thus, for all experiments, the success rates are above $98\%$.

\begin{figure*}
\includegraphics[width=0.8\textwidth]{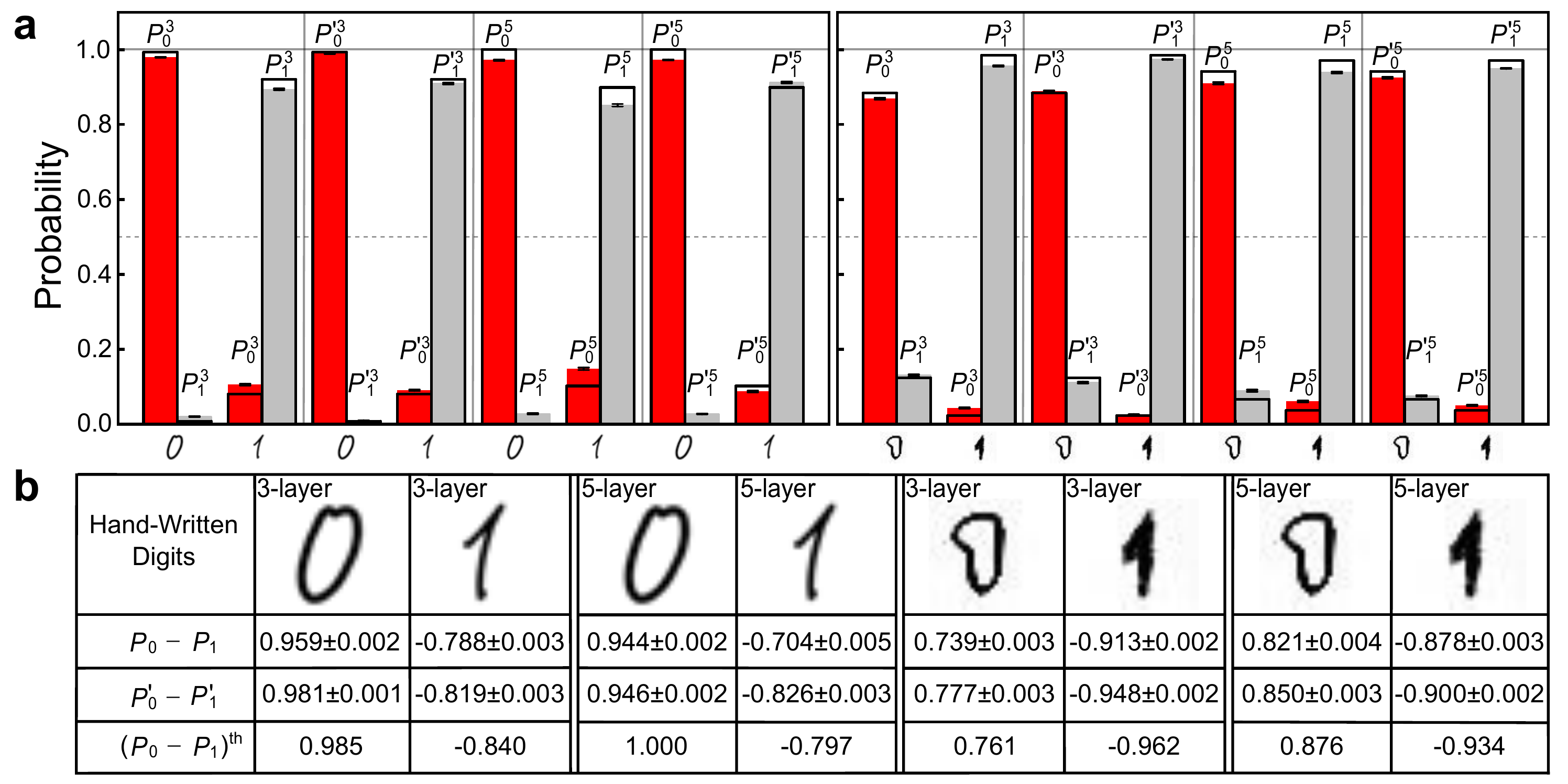}
\centering
\caption{Classification of images outside the MNIST dataset. (a) Measured probabilities of the output states of quantum classifiers. (b) Classification results.
}
\label{fig:fig5}
\end{figure*}

In Fig.~\ref{fig:fig5} we show the results of applying the trained classifier on two pairs of hand-written digits ``$0$'' and ``$1$'' that are not in the MNIST dataset. The first pair of digits are written in a standard way. The second pair is written such that the profile of ``$0$'' resembles that of ``$1$'' in the first pair, and the profile of ``$1$'' is much shorter and fatter compared to its counterpart in the first pair. For both cases, our classifier correctly recognizes the images with high confidence (large $|P_0-P_1|$), demonstrating the robustness and accuracy of the device.

%------------------------------------------------------------------------------------%

\section{Discussion}
\label{sec:conc}
We report the first experimental demonstration of quantum binary classification of real-life, hand-drawn images with single photons. The experimental scheme adopts a TN-based ML algorithm, which benefits from the powerful interpretability of TNs, as well as the efficient entanglement-based optimization in the quantum feature space.
After training and optimizing the classifier on classical computers, we use single photons to achieve the binary classification with {\red a} success rate over $98\%$.
Our experiment can be readily extended to multi-category classification, by taking advantage of the multiple degrees of freedom of photons and specially designed interferometric network for single- and two-qubit operations.
The hybrid quantum-classical scheme demonstrated here can be upgraded to be fully quantum mechanical, if the optimization in the ML process is performed on device parameters (such as the setting angles of wave plates), rather on the overall design of quantum circuits.
Furthermore, the TN-based ML algorithm is general, and directly applicable to a wide range of physical systems that have full control of qubits/qudits, including nitrogen-vacancy centers~\cite{LWL+19}, nuclear magnetic resonance systems~\cite{LLXD15SVMNMR}, and trapped ion~\cite{ZZW20QMLion}. In particular, with the rapid progress in quantum computers based on superconducting quantum circuits~\cite{AABB+2019GoogleQ}, it is hopeful that TN-based ML can be demonstrated for a larger feature space with more qubits, such that it can find utilities in more complicated real-world tasks.
%------------------------------------------------------------------------------------%

\begin{acknowledgements}
We thank Barry C. Sanders for providing constructive criticism on the manuscript.
This work has been supported by the Natural Science Foundation of China (Grant No. 12025401, No. 11674189, No. U1930402, and No. 11974331). K. W. and L. X.  acknowledge support from the Project Funded by China Postdoctoral Science Foundation (Grant Nos. 2019M660016 and 2020M680006). WY acknowledges support from the National Key Research and Development Program of China (Grant Nos. 2016YFA0301700 and 2017YFA0304100). SJR is supported by Beijing Natural Science Foundation (Grant No. 1192005 and No. Z180013) and in part by the NSFC (Grant No. 11834014), and by the Academy for Multidisciplinary Studies, Capital Normal University.
\end{acknowledgements}

\appendix

\renewcommand{\theequation}{A\arabic{equation}}
\setcounter{equation}{0}
\section{Tensor network machine learning algorithm}

For a classical gray-scale image consisting of $N$ features (pixels or frequency components), we follow the general recipe of TN-based supervised learning~\cite{SS16TNML,LRWP+17MLTN} and map the classical image to a product state of $N$ qubits in the quantum Hilbert space
\begin{align}\label{eq-prod_state}
|\phi \rangle = \bigotimes_{n=1}^N |s_n \rangle
\end{align}
with the feature map given by
\begin{align}\label{eq-map}
|s_n\rangle= \cos\frac{x_n\pi}{2} |0\rangle + \sin\frac{x_n\pi}{2} |1\rangle.
\end{align}
Here $0\leq x_n\leq 1$ characterizes the $n$th feature and determines the superposition coefficients of the $n$th qubit $|s_n\rangle$ in the basis $\{|0\rangle,|1\rangle\}$.

To classify a set of images into $N_c$ categories, we introduce a quantum classifier state $|\Psi\rangle$ in a joint Hilbert space $\mathcal{H}_\text{prod} \otimes \mathcal{H}_c$, where $\mathcal{H}_\text{prod}$ denotes the Hilbert space of the product state in Eq.~(\ref{eq-map}) and $\mathcal{H}_c$ is the $N_c$-dimensional Hilbert space encoding the information of different categories. The classifier state should be constructed in such a way that, for any given unclassified image with the mapped quantum state $|\phi\rangle$, the probability of this image belonging to the $c$th category is
\begin{align}
\label{eq-Pc}
P_c= \left| \langle \Psi | (|\phi \rangle \otimes |c\rangle)\right|^2,
\end{align}
where $\{|c \rangle\}$ (with $c = 0, \ldots, N_c-1$) represents the orthonormal basis in the Hilbert space $\mathcal{H}_c$. Hence, $P_c$ constitutes the probability distribution for different categories of a given classical image, which, as we demonstrate later, can be probed via projective measurements. The prediction of the classifier is given by the category with the largest probability (i.e., $\text{argmax}_c P_c$).

Following common practice in TN-based machine learning, we use the matrix product state (MPS)~\cite{PVWC07MPSRev,VMC08MPSPEPSRev} to represent the classifier state as
\begin{align}\label{eq-MPS}
|\Psi \rangle = \sum_{\{s\}} \sum_{\{a\}} \sum_{c} A^{[N]}_{s_N, c a_{N-1}} \ldots A^{[2]}_{s_{2}, a_{2} a_{1}}A^{[1]}_{s_1, a_{1}}
\bigotimes_{n=1}^N|i_n\rangle |c\rangle.
\end{align}
For the subscripts of tensors $A^{[i]}$ ($i=1,\ldots, N$),  $\{i_n\}$ are physical indices labeling feature qubits in the Hilbert space with $i_n=0$ or $1$. The virtual indices $\{a_n\}$ will be contracted in the simulation, and the label index $c$ is unique to $A^{[N]}$ in $\mathcal{H}_c$ and represents the categories.

For our purpose of using only two qubits, we consider binary classification problems with $\dim(c)=N_c=2$, and take the dimensions of the virtual indexes $\dim(a_n)=2$ ($n=1, \ldots, N-1$). The binary classifications exemplified in this work can be readily extended to digits other than ``$0$'' and ``$1$'', and to multi-category classifications with $N_c>2$, where higher-level qudits would be needed. With the MPS representation, the complexity of the state classifier scales only linearly with $N$, which enables us to efficiently optimize the classifier on classical computers.

Training the MPS on classical computers as in steps (i) and (iii) amounts to optimizing the tensors in Eq.~(\ref{eq-MPS}). To this end, we define the loss function as the negative logarithm of the probability of correctly categorizing all images in the training set
\begin{equation}\label{eq-lossf}
f = - \sum_{m \in \mathcal{I}}  \ln P_{c^{(m)}},
\end{equation}
where $\mathcal{I}$ denotes the training set, $P_{c^{(m)}}$ is defined by Eq.~(\ref{eq-Pc}) and $c^{(m)}$ denotes the correct category of the $m$th image. The training process thus involves the minimization of $f$ from the training set, which is equivalent to the maximization of the accuracy of classifying the training samples~\cite{B06MLbook}.

To translate the MPS into executable quantum circuits~\cite{SSVCW05PrepareMPS, HPWS18TNQML}, we apply the environment method~\cite{LRWP+17MLTN} in which tensors are isometries in the training process. Gate operations in the quantum circuit are then directly determined by the tensors $A^{[i]}$. Specifically, the tensors in the MPS satisfy the right-to-left orthogonal conditions
\begin{align}
\sum_{i_N a_{N-1}} A^{[N]}_{i_N, c a_{N-1}} A^{[N]}_{i_N, c' a_{N-1}} &= \one_{c c'}, \label{eq-ort1} \\
\sum_{i_n a_{n-1}} A^{[n]}_{i_n, a_{n} a_{n-1}} A^{[n]}_{i_n, a_{n}' a_{n-1}} &= \one_{a_{n} a'_{n}}, \label{eq-ort2} \\
\sum_{i_1} A^{[1]}_{i_1, a_1} A^{[1]}_{i_1, a_1'} &= \one_{a_1 a_1'}, \label{eq-ort3}
\end{align}
where $\one$ denotes the identity and $n=2, \ldots, N-1$. Under these orthogonal conditions, the MPS represents a renormalization-group flow from the right to the label index located at the left end.

%The orthogonal conditions allow us to encode the MPS into a quantum circuit that is executable on quantum systems~\cite{SSVCW05PrepareMPS, HPWS18TNQML}.

However, obviously the MPS given by Eq.~(\ref{eq-MPS}) contains as many qubits as the number of features in an image, which is far too large to implement on currently available quantum hardwares. Therefore, in step (ii), we reduce the required number of qubits by performing an entanglement-based optimization of the MPS architecture~\cite{LZLR18entTNML}. In essence, in the product state Eq.~(\ref{eq-MPS}), we only retain a handful of core feature qubits, denoted as $\{|\psi_i\rangle\}$ (note $\{|\psi_i\rangle\} \subseteq \{|s_i\rangle\}$), which possess the largest entanglement with other qubits in the classifier state $|\Psi\rangle$. The number of the extracted features $\tilde{N} = \# \{|\psi_i\rangle\}$ would be much smaller than $N$. Under a similar construction as Eq.~(\ref{eq-prod_state}), the product state for an image becomes $ \bigotimes_{i} |\psi_i\rangle$, and the number of tensors in the resulting MPS is significantly reduced from $N$ to $\tilde{N}$. Here, for any feature qubit $|s_n\rangle$, we characterize its bipartite entanglement using the entanglement entropy
\begin{equation}\label{eq:S}
	S^{[n]} =-\text{Tr} \hat{\rho}^{[n]} \ln \hat{\rho}^{[n]},
\end{equation}
where $\hat {\rho}^{[n]} = \text{Tr}_{/s_n} | \Psi\rangle \langle \Psi|$, and $\text{Tr}_{/s_n}$ traces over all degrees of freedom except $s_n$.

The entanglement-based optimization outlined above would work well, provided that the key information of an image should be carried by a small number of features. We ensure this by transforming the images into a data-sparse space using discrete-cosine transformation (DCT), such that the feature qubit $|s_n\rangle$ is constructed from frequency components rather than pixels. This is achieved as follows.

Consider a square, gray-scale image consisting of $N$ pixels, with the value of the pixel on the $i$th row and $j$th column characterized by  $x_{i,j}$ ($0\leq x_{i,j}\leq 1$).
To lower the required number of features for image classification, we transform the classical image data in the pixel space $\{x\}$ to the frequency space $\{y\}$ using a DCT
\begin{align}
& y_{p, q}=\frac{2}{H}\alpha(p-1)\alpha(q-1)\times \\ &\sum_{i=1}^{H}\sum_{j=1}^{H} x_{i, j}\cos{\left[\frac{(2i-1)(p-1)\pi}{2H}\right]}\cos{\left[\frac{(2j-1)(q-1)\pi}{2H}\right]}.\nonumber
\end{align}
Here $H$ is the height/width of the square image (in units of pixels), $p,q \in \{1, \ldots, H\}$, and $0\leq y_{p,q}\leq 1$. The factor $\alpha(p)=1/\sqrt{2}$ for $p=1$, and $\alpha(p)=1$ otherwise. In our case, we have $H=28$ for images in the MNIST dataset. The product state in Eq.~(\ref{eq-prod_state}) is therefore obtained by applying the feature map to the frequencies $\{y\}$.

To complete the feature extraction of the image, we retain a small number of feature qubits (three or five feature qubits corresponding to three- or five-layer constructions) that have the largest entanglement entropy in the quantum feature space, according to Eq.~(\ref{eq:S})~\cite{LZLR18entTNML}, to represent and classify hand-written digits of ``$0$'' and ``$1$'' in the MNIST dataset. It is crucial for implementing the classifier on our photonic simulator under three- or five-layer constructions, respectively. We note that since high-frequency components of the original image are mostly discarded in the cosine transformation, our scheme should also be robust to high-frequency noise in the hand-drawn image.

In our practical simulations of the ``0'' and ``1'' MPS classifier, the tensors in the MPS are initialized randomly. In consideration of efficiency, we randomly take 2000 images from the training set to train the MPS. The testing accuracy is evaluated by all the ``0'' and ``1'' digits in the testing set. The results can be reproduced by the code on \url{https://github.com/YuhanLiuSYSU/MPS_ImageClassifier}.

\renewcommand{\theequation}{B\arabic{equation}}
\setcounter{equation}{0}
\section{Qubit-efficient quantum circuits}

The MPS can be translated to a quantum circuit that consists of $\tilde{N}$ gates on $\tilde{N}$ qubits [step (iv)]. It follows that some matrix components of the gates are given directly by the tensors in the MPS as
\begin{align}
\langle 0c |U_N| i_N a_{N-1}\rangle &= A^{[N]}_{i_n, c a_{N-1}}, \label{eq-G1} \\
\langle 0 a_{n}|U_n |i_{n} a_{n-1} \rangle &= A^{[n]}_{i_n, a_{n} a_{n-1}}, \label{eq-G2} \\
\langle i_1|U_1| a_1\rangle &= A^{[1]}_{i_1, a_1}. \label{eq-G3}
\end{align}
Other components are determined by satisfying the following orthonormal conditions
\begin{align}
\langle 0c |U_N U_N^{\dagger} |1c'\rangle &= 0 \label{eq-G1o} \\
\langle 0 a_{n} | U_n U_n^{\dagger}| 1 a_{n}'\rangle &= 0, \label{eq-G2o}\\
\langle 1c |U_N U_N^{\dagger}| 1c'\rangle &= \one_{cc'}, \label{eq-G1n} \\
\langle 1 a_{n}| U_n U_n^{\dagger}| 1 a_{n}'\rangle &= \one_{a_{n} a_{n}'}. \label{eq-G2n}
\end{align}
After getting the matrix elements of all gate operations, we numerically determine the parameters of the photonic interferometry network.

The circuit still requires as many qubits as the retained features. To further reduce the number of qubits to two,
we adopt the quantum-efficient scheme~\cite{HPWS18TNQML} [step (v)], which is achieved by translating measurements on different qubits into those on only two qubits (the operational and classifier qubits) at different times. The extracted features of an image are input to the circuit by measuring the operational qubit at different times, while the information is carried by the classifier qubit, which can be extracted using projective measurements.

\renewcommand{\theequation}{C\arabic{equation}}
\setcounter{equation}{0}
\section{Equivalence between two schemes of quantum classifier}

\begin{figure*}%[b]
\centering
\includegraphics[width=0.8\textwidth]{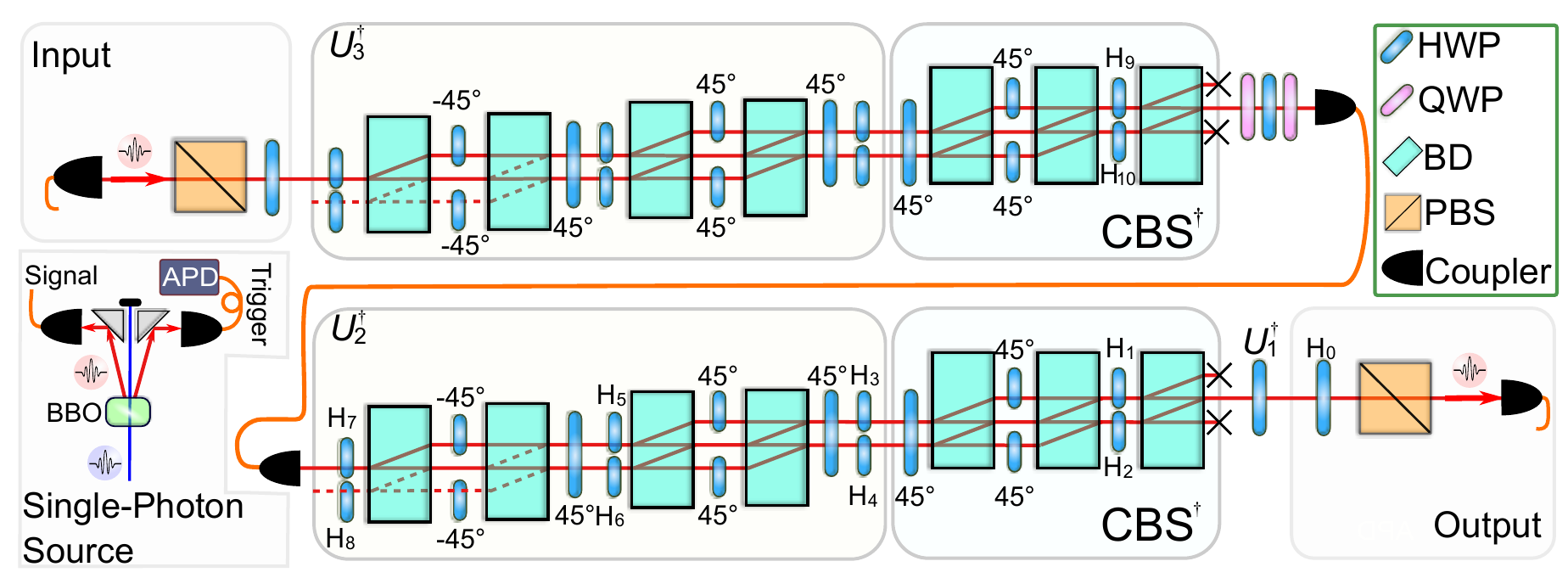}
\caption{Experimental demonstration of quantum classifier under Scheme B (three-layer construction). The classifier and operational qubits are initialized in $\ket{H}\otimes\ket{u}$ or $\ket{V}\otimes\ket{u}$ via a PBS and a HWP. After each gate operation $U^\dagger_i$ ($i=1,2,3$), a projective measurement $|\psi_i\rangle\langle \psi_i|$ is performed on the operational qubit via a CBS. This is the reverse process of the setup in Fig.~1 of the main text.
}
\label{fig:figS1}
\end{figure*}

The experimental setup for Scheme B is illustrated in Fig.~\ref{fig:figS1}, which is essentially the reverse process of Scheme A shown in Fig.~3 of the main text. Under Scheme B with the initial state $\ket{00}$, the probability $P'_0$ given by the last projective measurement can be written as (not normalized)
\begin{small}
\begin{align}
P'_0&=\text{Tr}\left[U_1^\dagger\rho_{20}U_1\ket{\psi_1}\bra{\psi_1}\right]\nonumber\\
&=\text{Tr}\left[\text{Tr}_{\rm op}\!\left(U^\dagger_2\!(\rho_{10}\!\otimes\!\ket{0}\!\bra{0})\!U_2\!(\one\!\otimes\!\ket{\psi_2}\!\bra{\psi_2})\right)\!U_1\!\ket{\psi_1}\!\bra{\psi_1}\!U^\dagger_1\right]\nonumber\\
&=\text{Tr}\left[\left(\rho_{10}\otimes\ket{0}\bra{0}\right)\tilde{\rho}\right],
\end{align}
\end{small}
where $\tilde{\rho}=U_2\left(U_1\ket{\psi_1}\bra{\psi_1}U^\dagger_1\otimes\ket{\psi_2}\bra{\psi_2}\right)U^\dagger_2$. Hence, $P'_0$ can be regarded as the joint probability of two local, projective measurements on $\tilde{\rho}$, with the outcome of the classifier qubit given by $\rho_{10}$ and that of the operational qubit given by $\ket{0}$. The probability can be re-arranged as
\begin{align}
P'_0&=\text{Tr}\left[\text{Tr}_{\rm op}\left(\tilde{\rho}(\one\otimes\ket{0}\bra{0})\right)\rho_{10}\right]\nonumber\\
&=\text{Tr}\left[\rho_1\text{Tr}_{\rm op}\left(U^\dagger_3\ket{00}\bra{00}U_3(\one\otimes\ket{\psi_3}\bra{\psi_3})\right)\right]\nonumber\\
&=\text{Tr}\left[U^\dagger_3\ket{00}\bra{00}U_3\left(\rho_1\otimes\ket{\psi_3}\bra{\psi_3}\right)\right]\nonumber\\
&=\text{Tr}\left[\ket{00}\bra{00}U_3\left(\rho_1\otimes\ket{\psi_3}\bra{\psi_3}\right)U^\dagger_3\right].
\end{align}

On the other hand, for Scheme A, the probability of the projective measurement $\sigma_z$ on the output classifier qubit is (not normalized)
\begin{align}
P_0&=\text{Tr}\left[\rho_2\ket{0}\bra{0}\right]\nonumber\\
&=\text{Tr}\left[\text{Tr}_{\rm op}\left(U_3(\rho_1\otimes\ket{\psi_3}\bra{\psi_3})U^\dagger_3(\one\otimes\ket{0}\bra{0})\right)\ket{0}\bra{0}\right]\nonumber\\
&=\text{Tr}\left[U_3\left(\rho_1\otimes\ket{\psi_3}\bra{\psi_3}\right)U^\dagger_3\ket{00}\bra{00}\right]=P'_0.
\end{align}
Similarly, as $P_0+P_1=P_0'+P_1'=1$, we have $P_1=P'_1$. Thus, we prove the equivalence of the two schemes.

\renewcommand{\theequation}{D\arabic{equation}}
\setcounter{equation}{0}
\section{Experimental details}

Experimentally, we create a pair of photons via spontaneous parametric down conversion, of which one serves as a trigger and the other serves as the signal~\cite{XZB+17}. The signal photon is then sent to the interferometry network. For both of our schemes, we encode the classifier qubit in the
polarization of the signal photon, with $|H\rangle$ and $|V\rangle$ corresponding to the horizontally and vertically polarized photons, respectively. The operational qubit is encoded in the spatial modes of the signal photon, with $|u\rangle$ and $|d\rangle$ representing the upper and lower spatial modes.

For all the 2115 images in the testing set, the mean success probability (the probability of obtaining the state after post-selections) is $92.1\%$ ($84.4\%$) for the three (five) layer constructions of the classifier. The efficiencies of the single-photon detectors are typically $66\%$. Furthermore, considering the losses caused by the optical elements and the coupling efficiencies, the typical success rate (defined as the ratio of detected to produced particles) of our setup is about $19.7\%$. For convenience, we tune the pump power to set the single-photon generation rate at about $2\times10^4$ per second. With the measurement time fixed at 3 s, we obtain the total photon count over $10^4$ for each image classification. Note that the measurement time can be drastically reduced if we upgrade the pump power of the single-photon source.

For both schemes, the single-qubit gate $U_1$ is realized by a HWP on the polarization of photons. In Scheme A, to prepare for the input of the two-qubit gates $U_i$ ($i=2,3$), the feature states, encoded in the spatial modes of photons as $\gamma|u\rangle+\eta|d\rangle$, are introduced by CBSs that expand the dimensions of the system: $\alpha|H\rangle+\beta|V\rangle\rightarrow(\alpha|H\rangle+\beta|V\rangle)\otimes(\gamma|u\rangle+\eta|d\rangle)$. A CBS is realized by three BDs and five HWPs. The first BD splits photons into different spatial modes depending on their polarizations. The following HWPs and BDs realize a controlled two-qubit gate on the polarizations and spatial modes of photons. Note that parameters of the unitary operators $U_i$ are fixed during the training process, which are encoded through the angles of HWPs.

The two-qubit gate $U_i$ is implemented by a cascaded interferometer. As an arbitrary $4\times 4$ matrix, $U_i$ can be decomposed using the cosine-sine decomposition method~\cite{I17}, where $U_2=LSR$, with $L$, $R$ and $S$ controlled two-qubit gates. $L$ and $R$ are realized by inserting the HWPs in the corresponding spatial mode, in which spatial mode serves as the control qubit and the polarization is the target qubit. For $S$, the polarization is the control qubit and the spatial mode is the target qubit. Thus, it can be further decomposed into a SWAP gate and a controlled gate, in which spatial mode is the control qubit and the polarization is the target one. These are realized by four BDs and several HWPs. In between each two-qubit gates, we use a $1$m long single-mode fiber to connect cascaded interferometers and act as a spatial filter.

\bibliography{classifier}

\end{document}